\documentclass[conference,letterpaper]{IEEEtran}

\usepackage{mathrsfs}
\usepackage[utf8]{inputenc} 
\usepackage[T1]{fontenc}
\usepackage{url}
\usepackage{ifthen}
\usepackage{cite}
\usepackage{amssymb}
\usepackage{graphicx}  \usepackage{float}
\usepackage[caption=false]{subfig}
\usepackage{stfloats}
\usepackage{algorithm}
\usepackage{algpseudocode}
\usepackage[cmex10]{amsmath}
\newtheorem{theorem}{Theorem}

\begin{document}
\title{Guessing What, Noise or Codeword?} 

\author{%
 \IEEEauthorblockN{Xiao Ma}
 \IEEEauthorblockA{School of Computer Science and Engineering\\
                   Guangdong Key Laboratory of Information Security Technology\\
                    Sun Yat-sen University\\
                  Guangzhou 510006, P. R. China\\
                   Email: maxiao@mail.sysu.edu.cn}
}



\maketitle


\begin{abstract}
In this paper, we distinguish two guessing algorithms for decoding binary linear codes. One is the guessing noise decoding~(GND) algorithm, and the other is the guessing codeword decoding~(GCD) algorithm. We prove that the GCD is a maximum likelihood~(ML) decoding algorithm and that the GCD is more efficient than GND for most practical applications. We also introduce several variants of ordered statistic decoding~(OSD) to trade off the complexity of the Gaussian elimination~(GE) and that of the guessing, which may find applications in decoding short block codes in the high signal-to-noise ratio~(SNR) region.
\end{abstract}


\section{Introduction}
It is well-known that maximum-likelihood~(ML) decoding is an NP-hard problem for a general linear block code~\cite{NPhard}.  However, it is feasible to implement ML decoding, especially when the soft information is available in the high signal-to-noise ratio~(SNR) region, for short codes which are crucial for ultra-reliable low-latency communication~(URLLC).

A typical example of exploiting the soft information is Chase decoding of block codes~\cite{chase1972class}, which repeatedly applies some decoding algorithm upon combinatorially flipping certain least reliable bits and selects the most likely one from the candidate codewords. This can be viewed as a guessing codeword decoding~(GCD) algorithm, which takes information set decoding~(ISD)~\cite{Prange1962} as an early example.
Another typical GCD, known as the ordered statistics decoding~(OSD)~\cite{OSD1995Fossorier}, produces a list of candidate codewords by re-encoding patterns in the most reliable basis~(MRB, an information set) with a small number of bits flipped. The OSD is universal and also near-optimal, which is applicable to any short linear block codes~(from low rates to high rates), including Bose-Chaudhuri-Hocquenghem~(BCH) codes, low-density parity-check~(LDPC) codes and polar codes,  resulting in capacity-approaching performance in the finite length region~\cite{Park2019,jiang2007reliability,wu2016ordered}. For the original OSD, the main computational complexity and the decoding latency are caused by the online Gaussian elimination (GE)  and the numerous re-encoding. The former can be circumvented for a BCH code by using Lagrange interpolation polynomials~\cite{ChenBCH2022} to form an extended systematic generator matrix for the corresponding Reed-Solomon~(RS) code~(not for the BCH code itself). While, the latter can be mitigated by, say, the segmentation-discarding OSD~(SD-OSD)~\cite{Yue2019SD}, the linear-equation OSD~(LE-OSD)~\cite{yue2022linear}, and the probability-based OSD~(PB-OSD)~\cite{Yue2021PB}.


In contrast to the GCD, the guessing random additive noise decoding~(GRAND) algorithm~\cite{GRAND2019} guesses the noise sequences from most likely to
least likely until the difference between the received vector and the guessing noise is a valid codeword. If the number of guessing is unlimited, the GRAND is definitely an ML algorithm, as also mentioned in the introductory paragraph of~\cite{GRANDidea}. The original GRAND has been generalized to, say, the soft-GRAND~(SGRAND)~\cite{sgrand}, GRAND with symbol reliability information~(SRGRAND)~\cite{SRGRAND}, and ordered reliability bits GRAND~(ORBGRAND)~\cite{ORBGRAND}. The GRAND-like algorithms are universal and can be applied to any codes~(linear or nonlinear), which do not rely on the code structure but require the code to have an efficient algorithm for membership checking. However, it is widely accepted that GRAND-like algorithms are only efficient~(in terms of complexity) for codes of short or moderate redundancy~\cite{ORBGRAND}.


In this paper, the GRAND-like algorithms are referred to as the guessing noise decoding~(GND) algorithm since they are also applicable to other noisy channels after transformation. We prove by analysis that the GCD is an ML decoding algorithm and that the GCD is more efficient than the GND in terms of the number of guessing. From a new perspective on the OSD as a special GCD, we summarize several variants of OSD, which trade off the complexity of GE and that of the re-encoding dominated by the number of guessing. Simulation results are provided to validate our analysis and show that the GCD requires a less number of guessing than the GND, especially for the low-rate codes.

\section{Problem Statement}
\subsection{System Model}
In this paper, we focus on applying binary linear block codes over discrete-time memoryless channels~(DMCs). Let $\mathbb{F}_2=\{0,1\}$ be the binary field and $\mathscr{C}[N,K]$ be a binary linear block code of dimension $K$ and length $N$. The binary linear block code $\mathscr{C}[N,K]$ can be specified either by a generator matrix $\mathbf{G}$ of size $K \times N$ or a parity-check matrix $\mathbf{H}$ of size $(N-K) \times N$. Associated with an information vector $\boldsymbol{u} \in \mathbb{F}_2^{K}$ is a codeword $\boldsymbol{c} = \boldsymbol{u} \mathbf{G}$, satisfying that $\mathbf{H}\boldsymbol{c}^{T}=\mathbf{0}$. Now suppose that $\boldsymbol{c} \in \mathbb{F}_2^{N}$ is transmitted over a DMC, resulting in  $\boldsymbol{y} \in \mathcal{Y}^N$, where $\mathcal{Y}$ is the alphabet of the channel outputs.

Upon receiving  $\boldsymbol{y}$, the log-likelihood ratio (LLR) vector $\boldsymbol{r}$ is calculated as
\begin{equation}\label{eq1}
    r_i = \log{\frac{P_{Y|C}(y_i|c_i=0)}{P_{Y|C}(y_i|c_i = 1)}},\ 0 \leq i < N,
\end{equation}
where $P_{Y|C}(\cdot|\cdot)$ is the conditional probability mass~(or density) function specifying the channel. Given the LLR vector $\boldsymbol{r}$, the hard-decision vector $\boldsymbol{z} \in \mathbb{F}_2^{N}$ is calculated as
\begin{equation}\label{harddecison}
	z_i=
	\begin{cases} 
		0, & \mbox{if }r_i\geq0\\
		1, & \mbox{if }r_i<0\\
	\end{cases}
	,~0\leq i < N.
\end{equation}
The ML decoding is to find a codeword $\boldsymbol{v}^*$ such that
\begin{equation}
 \boldsymbol{v}^*= \mathop{\text{argmax}}\limits_{\boldsymbol{v}\in \mathscr{C}} P_{Y|C}(\boldsymbol{y}|\boldsymbol{v}),
\end{equation}
which is equivalent to
\begin{equation}
 \boldsymbol{v}^*= \mathop{\text{argmin}}\limits_{\boldsymbol{v}\in \mathscr{C}} \log \frac{P_{Y|C}(\boldsymbol{y}|\boldsymbol{z})}{P_{Y|C}(\boldsymbol{y}|\boldsymbol{v})}.
\end{equation}
For a test vector $\boldsymbol{v} \in \mathbb{F}_2^{N}$, we can define its corresponding test error pattern~(TEP) $\boldsymbol{e}\in \mathbb{F}_2^N$ as
\begin{equation}
    \boldsymbol{e} \triangleq \boldsymbol{z}-\boldsymbol{v}.
\end{equation}
This can be written as $\boldsymbol{z}= \boldsymbol{v}+\boldsymbol{e}$ and hence the channel is transformed into an additive noise channel, which accepts the codeword as input and delivers the hard-decision vector as output. The distribution of additive noise can be time-varying, which depends on the original received vector $\boldsymbol{y}$ as well as the channel transition probability law. Defining the soft weight of a TEP $\boldsymbol{e}$, denoted by $\gamma(\boldsymbol{e})$, as
\begin{equation}
\gamma(\boldsymbol{e}) \triangleq \log  \frac{P_{Y|C}(\boldsymbol{y}|\boldsymbol{z})}{P_{Y|C}(\boldsymbol{y}|\boldsymbol{z}-\boldsymbol{e})} = \sum\limits_{i=1}^{N}e_i |r_i|,
\end{equation}
we see that the ML decoding is equivalent to the lightest-soft-weight decoding. That is,  the ML decoding is equivalent to
\begin{equation}
\begin{aligned}
    \min\limits_{\boldsymbol{e}\in \mathbb{F}_2^N} \quad &\gamma(\boldsymbol{e})\\
    \mbox{s.t.}\quad & \mathbf{H}\boldsymbol{e}^T=\boldsymbol{s}^T,
\end{aligned}
\end{equation}
where $\boldsymbol{s}^T=\mathbf{H}\boldsymbol{z}^T$ is the available syndrome.

\emph{Remark.} 
In the case when multiple valid TEPs are equally optimal, we assume that finding one of the lightest valid TEPs suffices to complete the decoding. For this reason, we assume in this paper that the lightest valid TEP is unique. By a valid TEP, we mean a TEP $\boldsymbol{e}$ that satisfies $\mathbf{H}\boldsymbol{e}^T = \boldsymbol{s}^T$.




\subsection{Guessing Noise Versus Guessing Codeword}
Without loss of generality, we assume that the first $N-K$ columns of $\mathbf{H}$ are linearly independent. That is, $\mathbf{H}$ can be transformed by elementary row operations into a systematic form,
\begin{equation}
    \mathbf{H} \rightarrow [\mathbf{I},\mathbf{P}],
\end{equation}
where $\mathbf{I}$ is the identity matrix of order $N-K$ and $\mathbf{P}$ is a matrix of size $(N-K) \times K$. Then a TEP can be written as $\boldsymbol{e} = (\boldsymbol{e}_\text{L},\boldsymbol{e}_\text{R})$, where $\boldsymbol{e}_L \in \mathbb{F}_2^{N-K}$ and $\boldsymbol{e}_{\text{R}} \in \mathbb{F}_2^K$. We see that, for any valid TEP $\boldsymbol{e}$, $\boldsymbol{e}_\text{L}$ is uniquely determined by $\boldsymbol{e}_\text{R}$ since $\boldsymbol{e}_\text{L}^T + \mathbf{P}\boldsymbol{e}_\text{R}^T = \boldsymbol{s}^T$.

We assume that a TEP sorter is available at the decoder that delivers $\boldsymbol{e}^{(i)}$ before $\boldsymbol{e}^{(j)}$ if $\gamma(\boldsymbol{e}^{(i)}) < \gamma(\boldsymbol{e}^{(j)})$ or $\gamma(\boldsymbol{e}^{(i)}) = \gamma(\boldsymbol{e}^{(j)})$ but $\boldsymbol{e}^{(i)}$ is prior to $\boldsymbol{e}^{(j)}$ in the lexicographic order. This can be implemented, say, with the aid of the flipping pattern tree~(FPT)~\cite{FPTtang}\cite{sgrand}\cite{Yue2021PB}\cite{LC_OSDljf2023}.  Then a sequence of TEPs $\boldsymbol{e} \in \mathbb{F}_2^N$ can be produced~(on demand) such that 
\begin{equation}\label{GNDSort}
\gamma(\boldsymbol{e}^{(0)}) \leq \gamma(\boldsymbol{e}^{(1)}) \leq \cdots \leq \gamma(\boldsymbol{e}^{(\ell)}) \leq \cdots \leq \gamma(\boldsymbol{e}^{(2^N-1)}).    
\end{equation}
Likewise, a sequence of partial TEPs $\boldsymbol{e}_{\text{R}} \in \mathbb{F}_2^K$ can be produced~(on demand) such that  
\begin{equation}\label{GCDSort}
 \gamma(\boldsymbol{e}_{\text{R}}^{(0)}) \leq \gamma(\boldsymbol{e}_{\text{R}}^{(1)}) \leq \cdots \leq \gamma(\boldsymbol{e}_{\text{R}}^{(\ell)}) \leq \cdots \leq \gamma(\boldsymbol{e}_{\text{R}}^{(2^K-1)}).   
\end{equation}

Given the sorted TEPs~\eqref{GNDSort}, a GND~\cite{GRAND2019,sgrand} is described in Algorithm~1. Likewise, given the sorted partial TEPs~\eqref{GCDSort}, a GCD is described in Algorithm~2. The differences between the GND and the GCD along with their complexity per guessing are analyzed below.
\begin{itemize}
    \item At the $\ell$-th guessing, the GND generates the $\ell$-th lightest TEP $\boldsymbol{e}^{(\ell)} \in \mathbb{F}_2^N$, while the GCD generates the $\ell$-th lightest partial TEP $\boldsymbol{e}_{\text{R}}^{(\ell)} \in \mathbb{F}_2^K$. The complexity is comparable for $K \approx N$.
    \item For the $\ell$-th TEP $\boldsymbol{e}^{(\ell)}$, the GNA calculates $\mathbf{H}\left(\boldsymbol{e}^{(\ell)}\right)^T $ for checking with a complexity of order $\mathcal{O}((N-K)N)$.  In contrast, the GND calculates  $\boldsymbol{e}_{\text{L}}^{(\ell)} = \boldsymbol{s}-\boldsymbol{e}_{\text{R}}^{(\ell)}\mathbf{P}^T$ with a complexity of order $\mathcal{O}((N-K)K)$, delivering a valid TEP  $\boldsymbol{e}^{(\ell)} = (\boldsymbol{e}_{\text{L}}^{(\ell)},\boldsymbol{e}_{\text{R}}^{(\ell)})$. Since the size of the matrix $\mathbf{P}$ is smaller than that of the matrix $\mathbf{H}$, the complexity of the re-encoding in the GCD is usually lower than the complexity of the checking in the GND unless $\mathbf{H}$ is a very sparse matrix but $\mathbf{P}$ is a dense matrix.
    \item  The checking in the GND compares $\mathbf{H}\boldsymbol{e}^T $ and $\boldsymbol{s}^T$, while the checking in the GCD compares  $\gamma(\boldsymbol{e}_{\text{R}})$ and $\gamma_{\text{opt}}$. The complexity is comparable.
    \item The GND checks TEPs with non-decreasing soft weights, but generates only one valid TEP at the final step.
    In contrast, the GCD re-encodes partial TEPs with non-decreasing soft weights, but generates multiple valid TEPs with $\gamma_{\text{opt}}$ decreases.
\end{itemize}

\emph{Remark.} It is worth pointing out that the GCD is different from the OSD since the GCD performs the GE offline, meaning that the GCD performs the GE only once, while the OSD usually requires to perform the GE for each reception of noisy codeword. Consequently, the complexity of transforming $\mathbf{H}$ into $[\mathbf{I}, \mathbf{P}]$ is not taken into account in the above analysis.

The total complexity can be roughly measured by the operations per guessing multiplied by the number of guessing. We have seen that the complexity of each guessing for the GCD is not higher than that of the GND. Then an immediate question arises: Can a GCD be
more efficient than a GND? The answer is positive, and the key is the early stopping criterion $\gamma(\boldsymbol{e}_{\text{R}})\geq \gamma(\boldsymbol{e}^*)$.

\begin{algorithm}[!t]
\renewcommand{\algorithmicrequire}{\textbf{Input:}}
    \renewcommand{\algorithmicensure}{\textbf{Output:}}
\caption{GND}\label{alg:alg1}
\begin{algorithmic}[1]
\Require The parity-check matrix $\mathbf{H}$, the LLR vector $\boldsymbol{r}$, and the available syndrome $\boldsymbol{s}$.
\State Initialization: $\ell = 0$, $\boldsymbol{e}^{(\ell)} = \boldsymbol{0}\in\mathbb{F}_2^N$.  
\While{$\mathbf{H} \left(\boldsymbol{e}^{(\ell)}\right)^{T} \neq \boldsymbol{s}^T$}
\State $\ell \gets \ell+1$.
\State Generate the $\ell$-th lightest TEP $\boldsymbol{e}^{(\ell)}$.
\EndWhile
\State $\boldsymbol{e}^* \gets \boldsymbol{e}^{(\ell)}$.
\Ensure The optimal searched codeword $\boldsymbol{c}^*= \boldsymbol{z}- \boldsymbol{e}^*$.
\end{algorithmic}
\label{alg1}
\end{algorithm}

\begin{algorithm}[t]
\renewcommand{\algorithmicrequire}{\textbf{Input:}}
    \renewcommand{\algorithmicensure}{\textbf{Output:}}
\caption{GCD}\label{alg:alg3}
\begin{algorithmic}[1]
\Require The parity-check matrix $[\mathbf{I},\mathbf{P}]$, the LLR vector $\boldsymbol{r}$,  and the available syndrome $\boldsymbol{s}$.
\State Initialization: $\ell = 0$, $\boldsymbol{e}_{\text{R}}^{(\ell)} = \boldsymbol{0}\in\mathbb{F}_2^K$, $\boldsymbol{e}_{\text{L}}^{(\ell)} = \boldsymbol{s}$.
\State $\boldsymbol{e}^{(\ell)} = (\boldsymbol{e}_{\text{L}}^{(\ell)} , \boldsymbol{e}_{\text{R}}^{(\ell)})$.
\State $\boldsymbol{e}_{\text{opt}} \gets \boldsymbol{e}^{(\ell)} $.
\State $\gamma_{\text{opt}} \gets \gamma(\boldsymbol{e}_{\text{opt}})$.
\While{$\gamma\left(\boldsymbol{e}_{\text{R}}^{(\ell)}\right) < \gamma_{\text{opt}}$ and $\ell < 2^K $}
\State $\ell \gets \ell+1$.
\State Generate the $\ell$-th lightest partial TEP  $\boldsymbol{e}_{\text{R}}^{(\ell)}$.
\If{$\gamma\left(\boldsymbol{e}_{\text{R}}^{(\ell)}\right) \geq \gamma_{\text{opt}}$}
\State \textbf{break}.
\Else
\State $\boldsymbol{e}_{\text{L}}^{(\ell)} = \boldsymbol{s}-\boldsymbol{e}_{\text{R}}^{(\ell)}\mathbf{P}^T$.
\State $\boldsymbol{e}^{(\ell)} = (\boldsymbol{e}_{\text{L}}^{(\ell)} , \boldsymbol{e}_{\text{R}}^{(\ell)})$.
\If{$\gamma\left(\boldsymbol{e}^{(\ell)}\right) < \gamma_{\text{opt}}$}
    \State $\boldsymbol{e}_{\text{opt}} \gets \boldsymbol{e}^{(\ell)}$.
    \State $\gamma_{\text{opt}} \gets \gamma\left(\boldsymbol{e}^{(\ell)}\right)$.
    \EndIf
\EndIf
\EndWhile
\Ensure The lightest TEP is $\boldsymbol{e}^* = \boldsymbol{e}_{\text{opt}}$, and the optimal searched codeword $\boldsymbol{c}^*= \boldsymbol{z}- \boldsymbol{e}^*$.
\end{algorithmic}
\label{alg3}
\end{algorithm}

\section{The Main Result}

\subsection{The Main Theorem}
\begin{theorem}
 The GCD is an ML algorithm, and the number of guessing for the GCD is less than or equal to the number of guessing for the GND.
  
\end{theorem}
\begin{IEEEproof}
The GCD terminates eventually. 
There are two cases when the GCD teminates. The first case is that the number of guessing reaches the maximum $2^K$. This occurs only when all partial TEP $\boldsymbol{e}_{\text{R}} \in \mathbb{F}_2^K$ are lighter than the lightest TEP. In this case, the GCD is equivalent to the exhaustive search, which is definitely an ML algorithm and $\boldsymbol{e}_{\text{opt}}$ is the lightest TEP. The second case is that $ \gamma(\boldsymbol{e}_{\text{R}}^{(\ell)}) \geq \gamma(\boldsymbol{e}_{\text{opt}})$. In this case, we have $\gamma({\boldsymbol{e}_{\text{opt}}}) \leq \gamma(\boldsymbol{e}_{\text{R}}^{(\ell)}) \leq \gamma(\boldsymbol{e}_{\text{R}}^{(j)})$ for all $j > \ell$. This implies that $\gamma(\boldsymbol{e}_{\text{opt}}) \leq \gamma(\boldsymbol{e})$ for all unexplored valid TEPs $\boldsymbol{e}$ since $\gamma(\boldsymbol{e}) = \gamma(\boldsymbol{e}_{\text{L}}) + \gamma(\boldsymbol{e}_{\text{R}})$, suggesting that $\boldsymbol{e}_{\text{opt}}$ is the lightest~(valid) TEP and further searches are not necessary.

Now assume that $\boldsymbol{e}^* = (\boldsymbol{e}_{\text{L}}^*,\boldsymbol{e}_{\text{R}}^*)$ is the lightest TEP, which is not known in advance but exists. The GND terminates eventually, and checks a list $\mathcal{L}_{\text{GND}} = \mathcal{S}_{\text{GND}} \cup \mathcal{T}_{\text{GND}}$, where $\mathcal{S}_{\text{GND}} = \{\boldsymbol{e} \in \mathbb{F}_2^N \ |\  \gamma(\boldsymbol{e}) < \gamma(\boldsymbol{e}^*) \}$ and $\mathcal{T}_{\text{GND}}$ is a subset of $\{\boldsymbol{e}\in \mathbb{F}_2^N \ |\  \gamma(\boldsymbol{e})=\gamma(\boldsymbol{e}^*)\}$.  In contrast, the GCD terminates with $\boldsymbol{e}_{\text{opt}}=\boldsymbol{e}^*$ and re-encodes a list 
$\mathcal{L}_{\text{GCD}} = \mathcal{S}_{\text{GCD}} \cup \mathcal{T}_{\text{GCD}}$, where $\mathcal{S}_{\text{GCD}}=\{\boldsymbol{e}_{\text{R}}\in \mathbb{F}_2^K \ |\ \gamma(\boldsymbol{e}_{\text{R}}) < \gamma(\boldsymbol{e}^*) \}$ and $\mathcal{T}_{\text{GCD}}=\{\boldsymbol{e}_{\text{R}} \in \mathbb{F}_2^K\ |\  \gamma(\boldsymbol{e}_{\text{R}})  = \gamma(\boldsymbol{e}^*) \}$. The set $\mathcal{T}_{\text{GND}}$~(if non-empty) consists of those (invalid) TEPs  $\boldsymbol{e}\in \mathbb{F}_2^N$  that  satisfy  $\gamma(\boldsymbol{e})=\gamma(\boldsymbol{e}^*)$ but are prior to $\boldsymbol{e}^*$ in the lexicographic order. In contrast, the set $\mathcal{T}_\text{GCD}$~(if non-empty) consists of those partial TEPs $\boldsymbol{e}_{\text{R}}\in \mathbb{F}_2^K$ that satisfy $\gamma(\boldsymbol{e}_{\text{R}})=\gamma(\boldsymbol{e}^*)$ but are prior to $\boldsymbol{e}_{\text{R}}^*$ in the lexicographic order. This occurs only when $\gamma(\boldsymbol{e}_{\text{L}}^*) = 0$ and hence $\gamma(\boldsymbol{e}_{\text{R}}) = \gamma(\boldsymbol{e}_{\text{R}}^*) = \gamma(\boldsymbol{e}^*)$ since, otherwise, $\gamma(\boldsymbol{e}_{\text{R}}) \leq \gamma(\boldsymbol{e}_{\text{R}}^*) < \gamma(\boldsymbol{e}^*)$.



For any $\boldsymbol{e}_{\text{R}}\in \mathcal{L}_{\text{GCD}}$, we construct a TEP $\boldsymbol{e}=(\boldsymbol{0},\boldsymbol{e}_{\text{R}})$ with $\boldsymbol{0}\in \mathbb{F}_2^{N-K}$. We have $\boldsymbol{e} \in \mathcal{L}_{\text{GND}}$ since either $\gamma(\boldsymbol{e}) < \gamma(\boldsymbol{e}^*)$ or $\gamma(\boldsymbol{e}) = \gamma(\boldsymbol{e}^*)$ but is prior to $\boldsymbol{e}^*$ in the lexicographic order. The latter case is true since $\boldsymbol{e}_{\text{R}}$ is prior to $\boldsymbol{e}_{\text{R}}^*$ and hence $\boldsymbol{e}$ is prior to $\boldsymbol{e}^*$ in the lexicographic order. Thus we have constructed an injective mapping $\boldsymbol{e}_{\text{R}} \rightarrow \boldsymbol{e}=(\boldsymbol{0},\boldsymbol{e}_{\text{R}})$ from $\mathcal{L}_{\text{GCD}}$ into $\mathcal{L}_{\text{GND}}$. This completes the proof that $|\mathcal{L}_{\text{GCD}}| \leq |\mathcal{L}_{\text{GND}}|$.

\end{IEEEproof}

\subsection{Illustrative Examples}
\textbf{Example 1}~(A toy example)\textbf{:}
Consider the Hamming code $\mathscr{C}[7,4]$ over a binary symmetric channel~(BSC) with cross error probability $p < 1/2$. No matter what codeword is transmitted and what vector is received, the GND will find the lightest TEP $\boldsymbol{e}^*$ with at most 8 guesses, one for the all zero TEP and 7 for the TEPs with Hamming weight one. The first guess is successful if and only if the true error pattern is a codeword, which occurs with a probability $p_0 = (1-p)^7 + 7p^3(1-p)^3 + p^7$. Hence, the average number of guessing for the GND is given by $p_0 + 35p_1$ with $p_1 = (1-p_0)/7$. In contrast, the maximum number of guessing for the GCD is 5, one for the all-zero partial TEP and 4 for the partial TEPs with Hamming weight one. The first guess is successful if and only the TEP~(obtained from $\boldsymbol{e}_{\text{R}}=\boldsymbol{0}$ by re-encoding) has a Hamming weight zero or one. In either case, further guessing is not necessary because all the remaining guesses are for $\boldsymbol{e}_{\text{R}}$ with $W_H(\boldsymbol{e}_{\text{R}}) \geq 1$ and must deliver $\boldsymbol{e}$ with $W_H(\boldsymbol{e}) \geq W_H(\boldsymbol{e}_{\text{R}}) \geq 1$. The probability that the first guess is successful is given by $p_0 + 3p_1$. The average number of guessing for the GCD is given by $p_0 + 17p_1$, which is strictly less than the average number of guessing for the GND. 




\textbf{Example 2:}
Consider a binary linear block code $\mathscr{C}[N,K]$ over a BSC. In this case, the soft weight is equivalent to Hamming weight. Suppose that $\boldsymbol{e}^*=(\boldsymbol{e}_{\text{L}}^*,\boldsymbol{e}_{\text{R}}^*)$ is the lightest valid TEP, which is not known in advance but exists. There are two cases. One is $W_H(\boldsymbol{e}_{\text{L}}^*) >0 $ and the other is $W_H(\boldsymbol{e}_{\text{L}}^*) =0 $.
 \begin{itemize}
        \item  For $W_H(\boldsymbol{e}_{\text{L}}^*) >0 $, the GCD will definitely find $\boldsymbol{e}_{\text{opt}}=\boldsymbol{e}^*$ within $\sum_{i=0}^{W_H(\boldsymbol{e}_{\text{R}}^*)}\binom{K}{i} $ guesses. The GCD continues the search since it cannot check whether $\boldsymbol{e}_{\text{opt}}$ is the lightest one or not. As the search proceeds, $W_H(\boldsymbol{e}_{\text{R}})$ increases but $\boldsymbol{e}_{\text{opt}}\ (=\boldsymbol{e}^*)$ keeps unchanged. Once all $\boldsymbol{e}_{\text{R}} \in \mathbb{F}_2^K$ with $W_H(\boldsymbol{e}_{\text{R}}) < W_H(\boldsymbol{e}^*)$ have been re-encoded, the GCD can safely confirm that $\boldsymbol{e}_{\text{opt}}$ is the lightest TEP. Therefore, the total number of guessing for the GCD is $\min\left\{2^K, \sum_{i=0}^{W_H(\boldsymbol{e}^*)-1}\binom{K}{i}\right\}$, which is strictly less than $\sum_{i=0}^{W_H(\boldsymbol{e}^*)-1}\binom{N}{i}$, a lower bound on the number of guessing for the GND.
        \item  For $W_H(\boldsymbol{e}_{\text{L}}^*)=0$, the number of guessing for the GCD is $\sum_{i=0}^{W_H(\boldsymbol{e}^*)-1}\binom{K}{i} + T$, where $T$ is the rank of $\boldsymbol{e}_{\text{R}}^*$~(according to the lexicographic order)    
        in the set $\{\boldsymbol{e}_{\text{R}}\in \mathbb{F}_2^K| W_H(\boldsymbol{e}_{\text{R}}) = W_H(\boldsymbol{e}_{\text{R}}^*)\}$. Again, this number is strictly less than $\sum_{i=0}^{W_H(\boldsymbol{e}^*)-1}\binom{N}{i} + T'$, the number of guessing for the GND, where $T'$ is the rank of $\boldsymbol{e}^*$ in the set $\{\boldsymbol{e}\in \mathbb{F}_2^N| W_H(\boldsymbol{e}) = W_H(\boldsymbol{e}^*)\}$.
    \end{itemize}

\emph{Remark.} Notice that, for high-rate codes, $N\approx K$ and $\binom{N}{i} \approx \binom{K}{i}$. The excess search number for GND over the GCD can be small.

\begin{figure}[!t]
\centering
\includegraphics[width=2.8in]{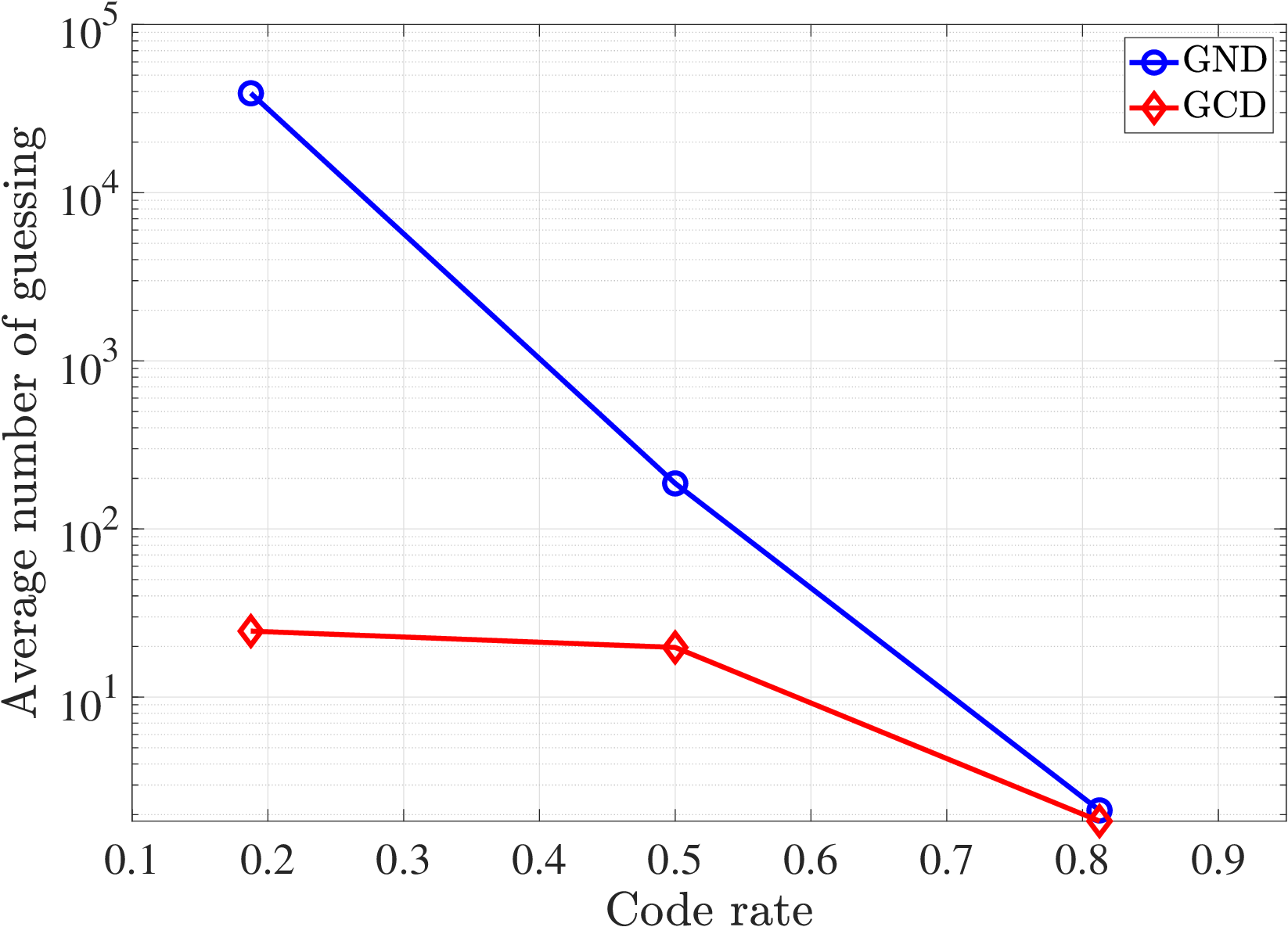}
\caption{Average number of guessing for the GND and the GCD. The simulations are conducted for three RM codes over the BPSK-AWGN channels at the target $\text{FER}=10^{-3}$.}
\label{Query}
\end{figure}

\textbf{Example 3:}
Consider three Reed-Muller~(RM) codes, $\mathscr{C}_{\text{RM}}[32,6]$, $\mathscr{C}_{\text{RM}}[32,16]$ and $\mathscr{C}_{\text{RM}}[32,26]$, over an additive white Gaussian noise channel~(AWGN) with binary phase shift keying~(BPSK) modulation. Shown in Fig.~\ref{Query} are the average numbers of guessing per reception of noisy codeword for the GCD and the GND at target frame error rate~(FER) $10^{-3}$~(corresponding to different SNRs for different code rates). We see that the GCD requires fewer guesses than the GND, validating our analysis. We also see that the  gap between the number of guessing is narrowed as the code rate increases. This suggests that, compared with the GND, the GCD is more universal and applicable to codes with a wide range of code rates.  

\section{OSD and Its Variants}
\subsection{A New Perspective on OSD}
As a special GCD, the OSD requires sorting to find the MRB, whose motivation can be understood from another perspective. First, the sorting makes $\gamma(\boldsymbol{e}_{\text{R}}^*)$ as small as possible, and hence the lightest one can enter into the list as early as possible. Second, the sorting makes $\gamma(\boldsymbol{e}_{\text{R}})$ increases as fast as possible to guarantee the earlier termination with the condition $\gamma(\boldsymbol{e}_{\text{R}}) \geq \gamma(\boldsymbol{e}^*)$. The cost of the OSD is the online GE for each noisy reception, and the benefit of the OSD over a general GCD is the reduction of the number of guessing.

\subsection{Variants of OSD}
For the conventional OSD, two issues arise: 1) how to skip those
unpromising TEPs; and 2) how to reduce the complexity of the
online GE. As a special class of GCD, several variants of the OSD have been proposed to address these issues, as summarized below.

\subsubsection{LC-OSD~\rm{\cite{LC_OSD2022,LC_OSDljf2023,LC_OSDTIT}}}

The objective of the LC-OSD is to reduce the number of guessing. The is achieved by introducing an extended MRB of size $K+\delta$ and searching partial TEPs over a trellis with the serial list
Viterbi algorithm~(SLVA)~\cite{SLVA}. By so doing, many unnecessary TEPs are skipped, and the soft weight $\gamma(\boldsymbol{e}_{\text{R}})$ increases faster.

\subsubsection{Representative OSD~(ROSD)~\rm{\cite{ROSD}}}
The ROSD is to reduce the complexity of the GE, which is applicable to a class of codes, called staircase generator matrix codes.

\subsubsection{Quasi-OSD~\rm{\cite{QuasiOSD}}}
Another way to reduce the delay caused by the GE is to relax the requirement of the MRB. In other words, we may perform the GCD with a relatively reliable basis instead of the MRB. With this relaxation, the GE can be replaced by, say, Lagrange interpolation for the (shortened) RS codes.  

\subsection{Simulation Results}
\textbf{Example 4:}
Consider an extended BCH~(eBCH) code $\mathscr{C}_{\text{eBCH}}[128, 64]$ over a BPSK-AWGN channel. 
The simulation results for the OSD algorithm and the LC-OSD algorithm are presented in Fig.~2, from which we observed that the LC-OSD performs similarly to the OSD but requires a much less average number of TEPs.

\begin{figure}[t]
  \centering
  \subfloat[Performance.\label{fig:bch-fer}]{\includegraphics[width=2.47in]{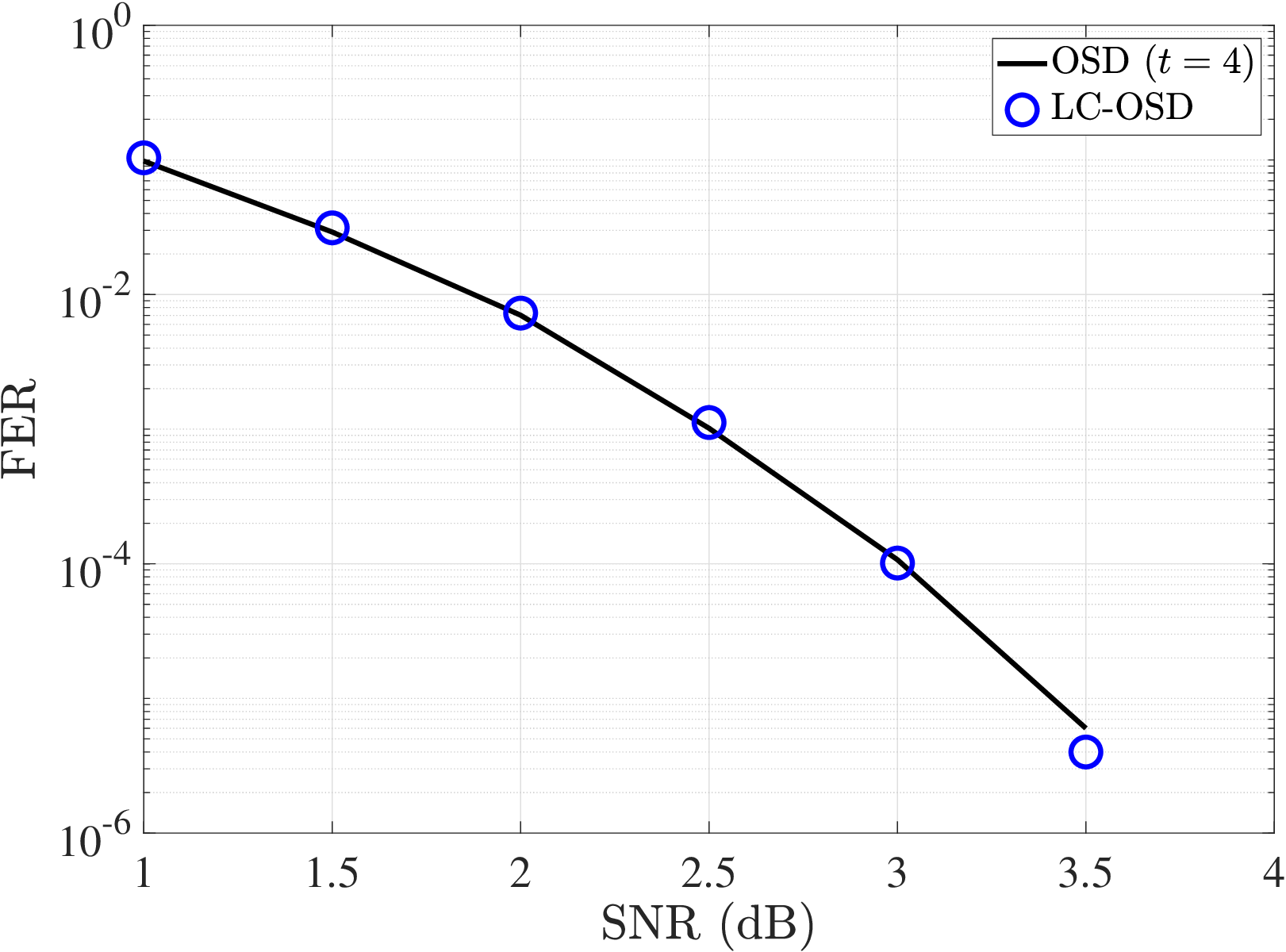}} 
  \vspace{0.1cm}
  \subfloat[Average number of TEPs.\label{fig:bch-timing}]{\includegraphics[width=2.47in]{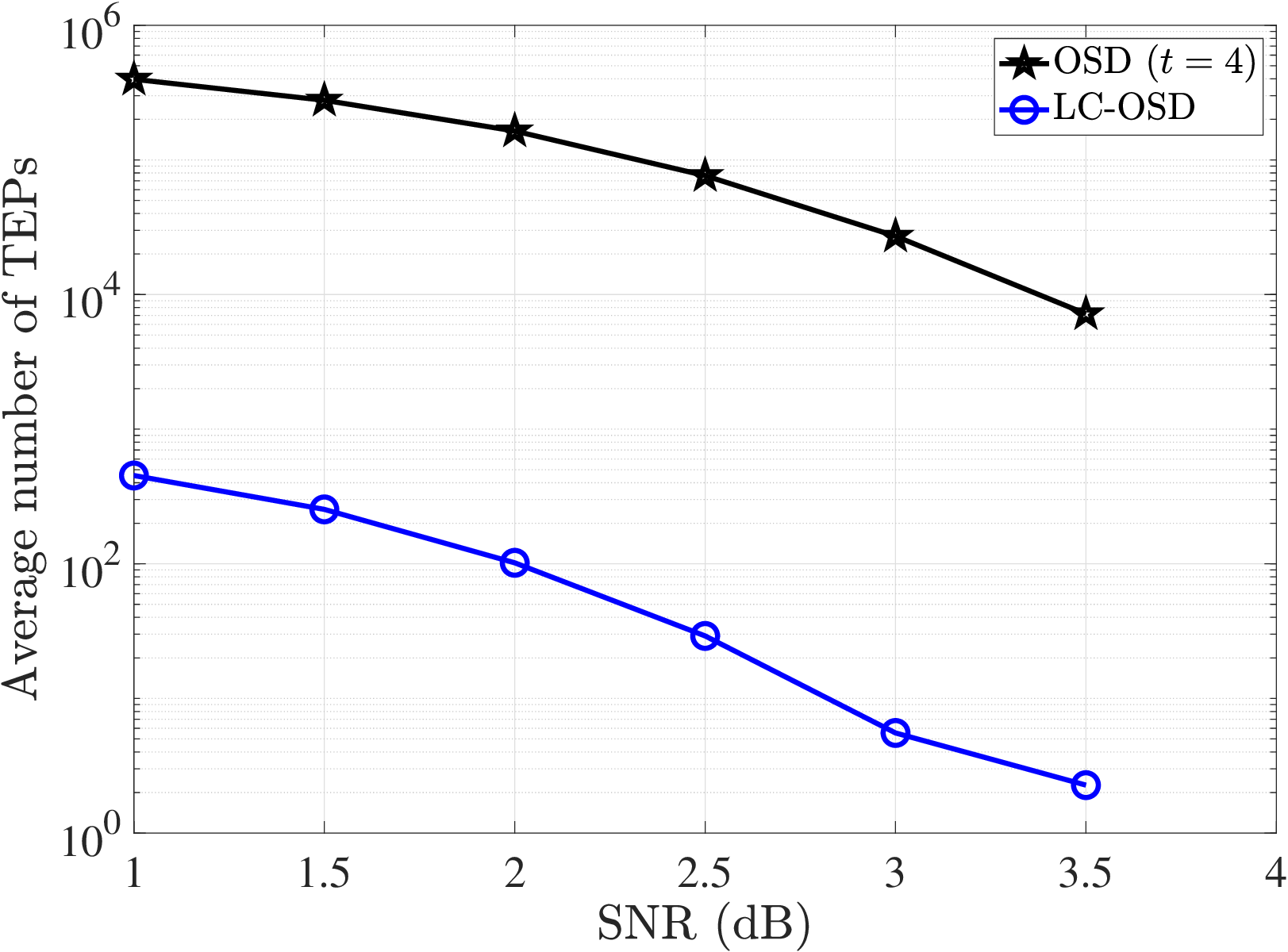}} 
  \caption{
    Simulation results of the eBCH code $\mathscr{C}_{\text{eBCH}}[128,64]$.
    Here, the maximum number of TEPs $\ell_{\text{max}}=2^{14}$ and $\delta = 8$ for LC-OSD.
  }
  \label{fig:bch}
\end{figure}

\textbf{Example 5:}
Consider an RM code $\mathscr{C}_{\text{RM}}[128,64]$ over a BPSK-AWGN channel. The simulation results for the LC-OSD algorithm and the ROSD algorithm are presented in Fig.~3, from which we observed that the performance of ROSD is similar to that of the  LC-OSD. The complexity of the GE for ROSD is reduced but at the cost of increase in the average number of TEPs.

\begin{figure}[!t]
  \centering
  \subfloat[Performance.\label{fig:bch-fer}]{\includegraphics[width=2.47in]{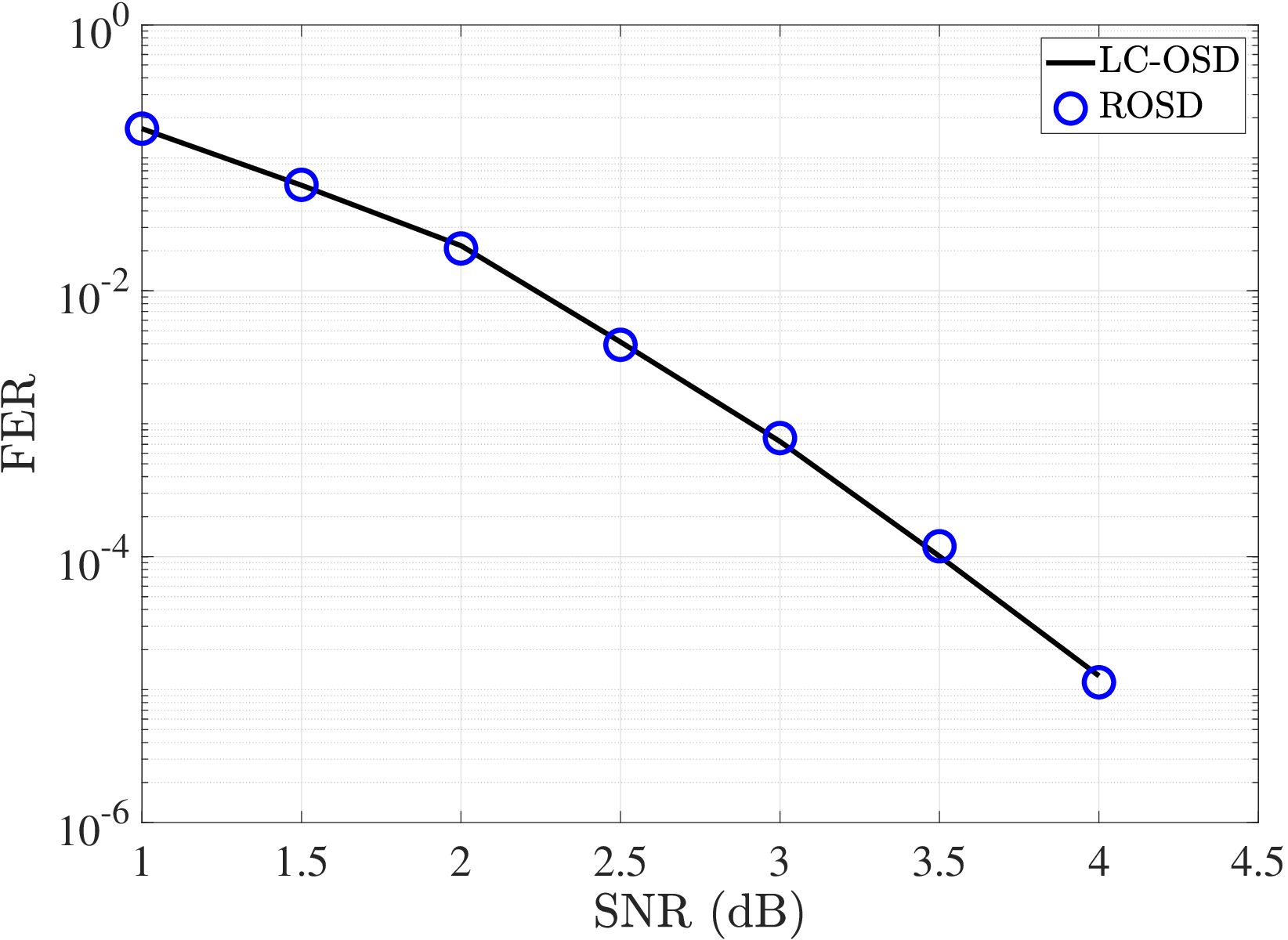}} \vspace{0.1cm}
  \subfloat[Average number of TEPs.\label{fig:bch-timing}]{\includegraphics[width=2.47in]{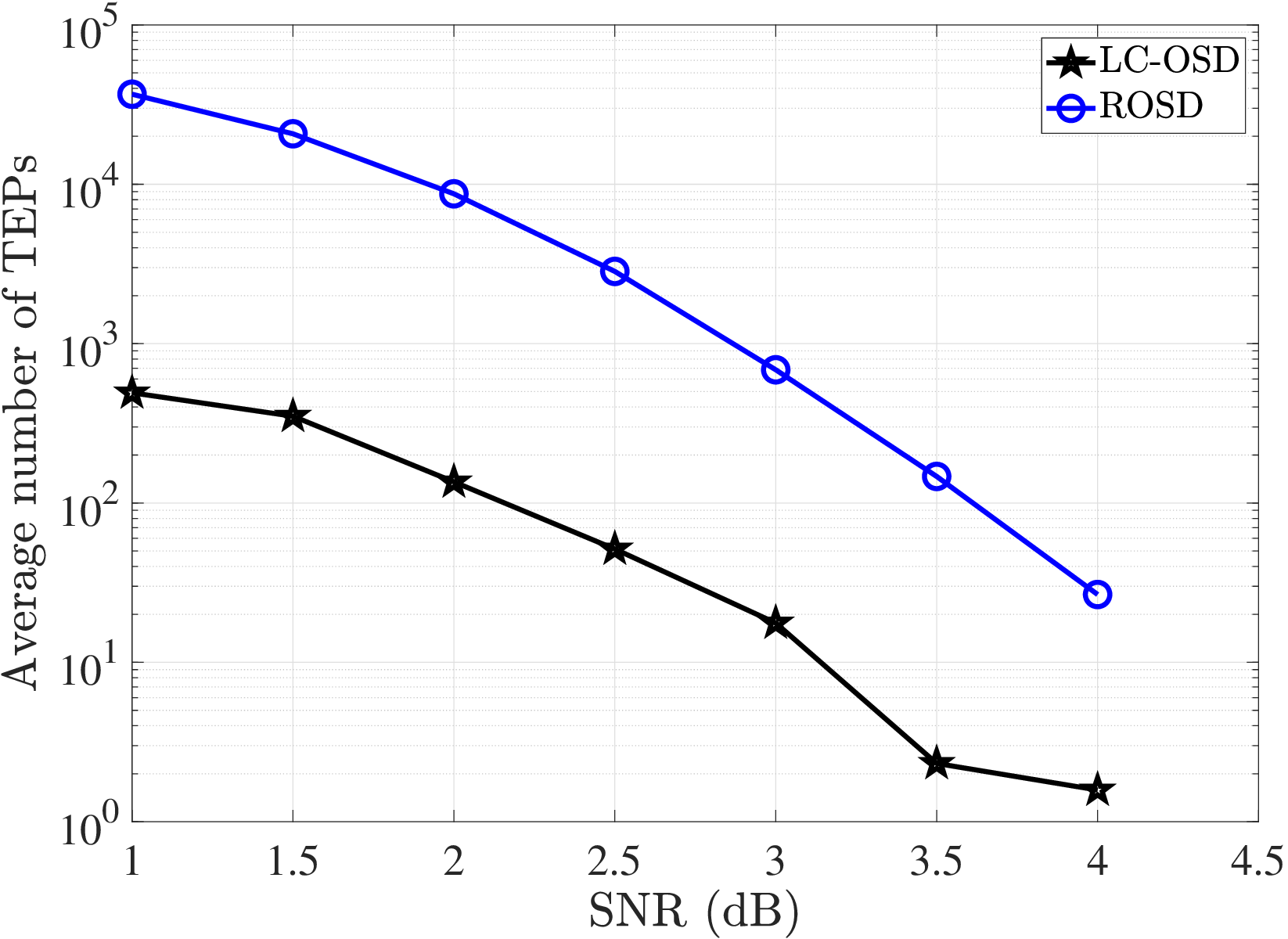}} 
  \caption{
    Simulation results of the RM code $\mathscr{C}_{\text{RM}}[128,64]$.
    Here, the maximum number of TEPs $\ell_{\text{max}}=10^6$ and $\delta = 12$ for LC-OSD and ROSD.
  }
  \label{fig:bch}
\end{figure}

\textbf{Example 6:}
Consider a shortened RS code $\mathscr{C}_{\text{RS}}[26,23]_{2^5}$ which is defined over the field $\mathbb{F}_{2^5}$ and mapped into $\mathbb{F}_2^{130}$, over a BPSK-AWGN channel. The simulation results are presented in Fig.~4, from which we observed that the quasi-OSD performs similarly to the GND but requires a much less average number of TEPs. We also observed that, compared with the GND, the GCD~(without online GE) requires a less number of TEPs~(on average) than the GND. The reduction in the number of guessing is not significant since the code rate is high.


\begin{figure}[!t]
  \centering
\subfloat[Performance.\label{fig:bch-fer}]{\includegraphics[width=2.47in]{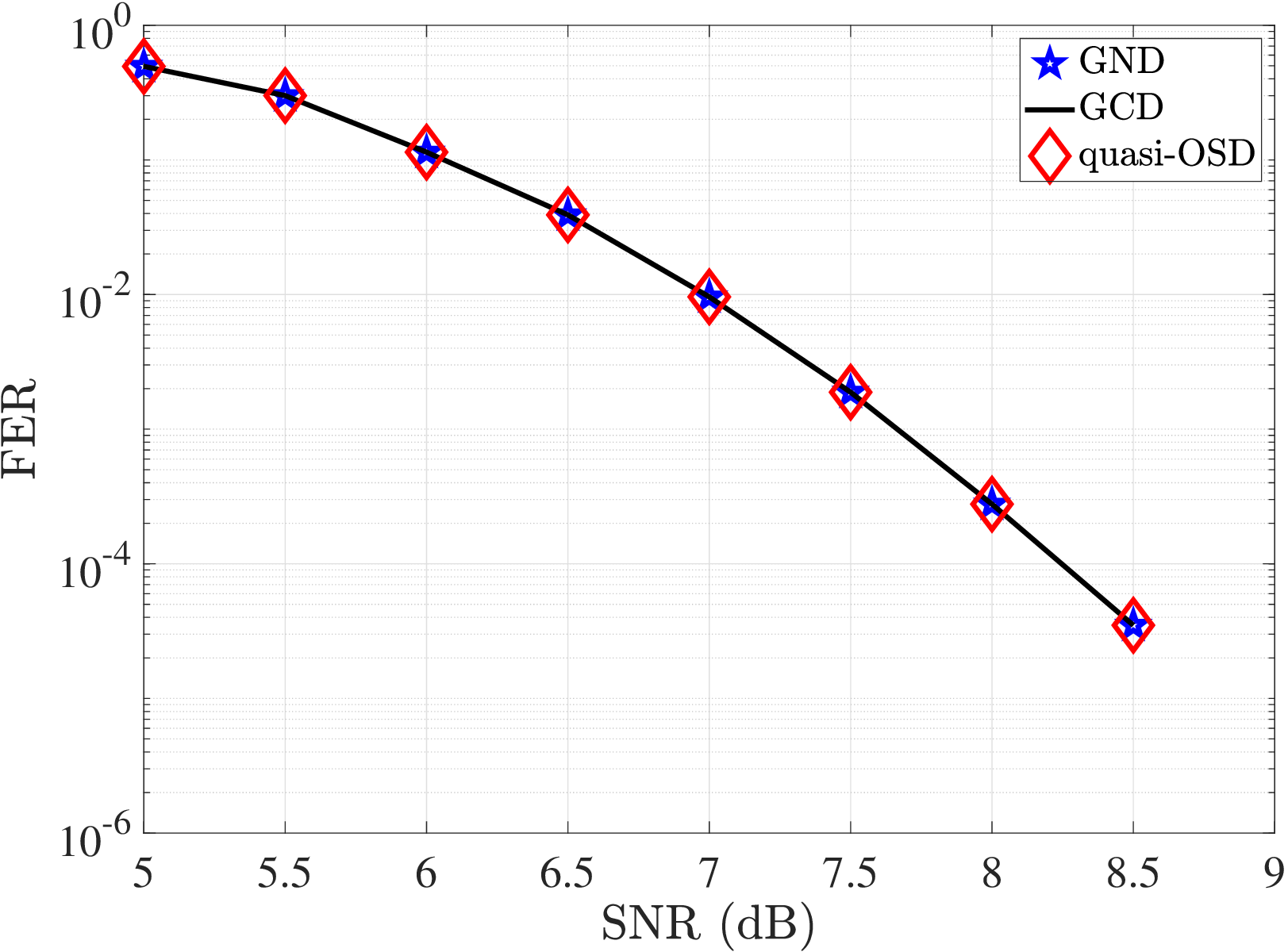}} 
  \vspace{0.1cm}
  \subfloat[Average number of TEPs.\label{fig:bch-timing}]{\includegraphics[width=2.47in]{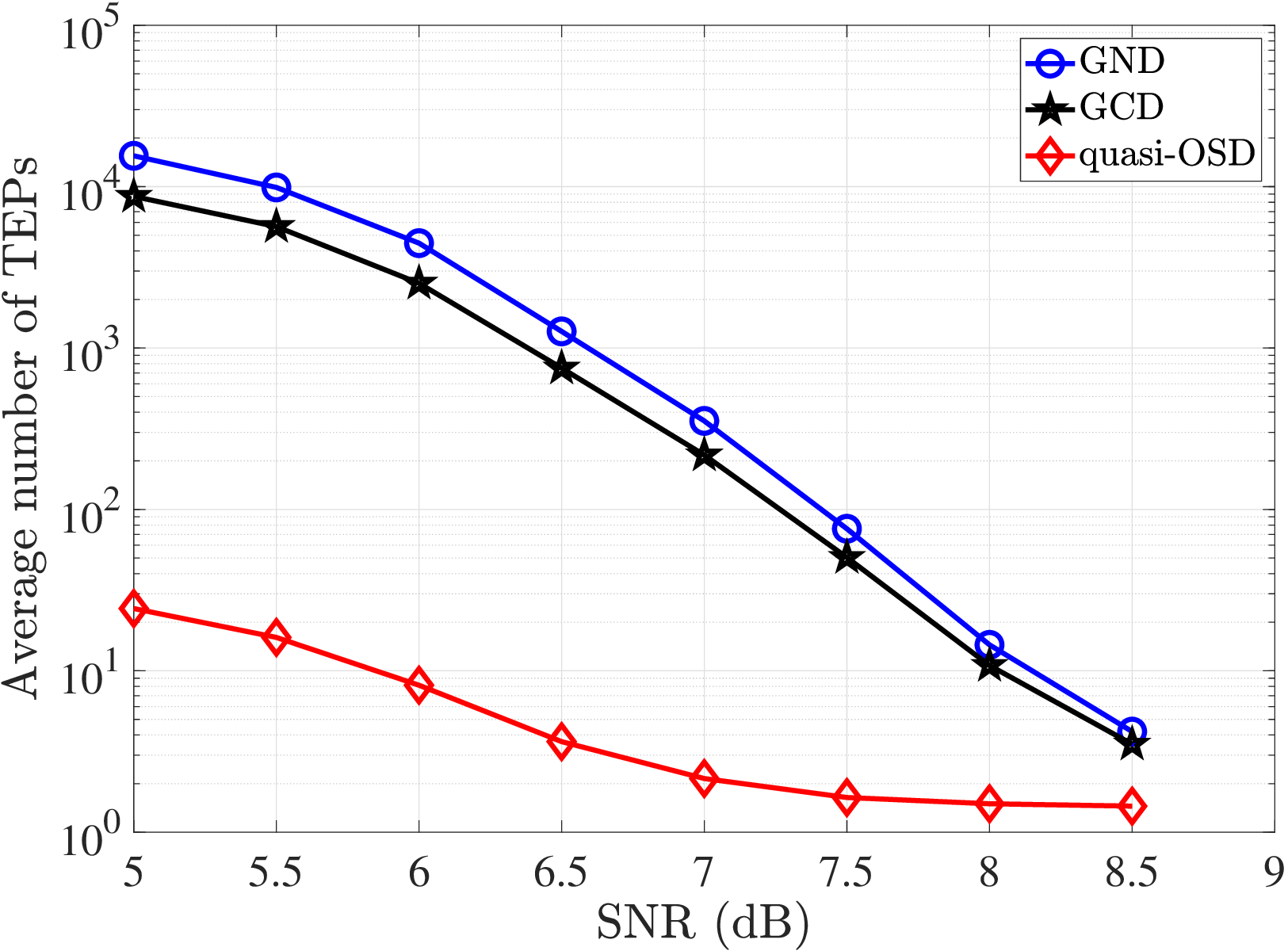}} 
  \caption{
    Simulation results of the shortened RS code $\mathscr{C}_{\text{RS}}[26,23]_{2^5}$.
    Here, the maximum number of TEPs $\ell_{\text{max}}=10^6$ for quasi-OSD, GCD and GND, and $\delta = 8$ for quasi-OSD.
  }
  \label{fig:bch}
\end{figure}

\section{Conclusion}
We have presented a general GCD algorithm for binary linear codes, which does not require online GE. We prove that the GCD is an ML algorithm and requires a less or equal number of guessing than the GND. As a final remark, we want to emphasize that no decoding algorithm is universally optimal~(in terms of complexity). Taking OSD as an example, the guessing dominates the complexity in the low SNR region, while the GE dominates the complexity in the high SNR region. Depending on SNRs, we may perform online GE for MRB, online reduced-complexity GE for quasi-MRB, or offline GE, leading to several variants of OSD, some of which are designed for reducing the guessing number, while some are for reducing the complexity and delay caused by the GE.




\bibliographystyle{IEEEtran}
\bibliography{ref}

\end{document}